\documentclass[conference]{IEEEtran}
\usepackage{times}
\usepackage{helvet}
\usepackage{courier}

\usepackage{float}
\usepackage{subfig}
\usepackage{graphicx}
\usepackage{amsmath}
\usepackage{multirow}
\usepackage{booktabs}
\usepackage{paralist}
\usepackage{url}
\usepackage{color}
\usepackage{cite}
\usepackage{epstopdf}
\usepackage[justification=centering]{caption}
\usepackage{tabularx}
\usepackage{algorithm}
\usepackage{algorithmic} 
\newcolumntype{L}[1]{>{\raggedright\let\newline\\\arraybackslash\hspace{0pt}}m{#1}}
\newcolumntype{C}[1]{>{\centering\let\newline\\\arraybackslash\hspace{0pt}}m{#1}}
\newcolumntype{R}[1]{>{\raggedleft\let\newline\\\arraybackslash\hspace{0pt}}m{#1}}
\newcommand{\name}{}
\def\name/{MLP}

\ifCLASSINFOpdf

\else

\fi

\begin{document}

%
\title{A Holistic Approach for Predicting Links in Coevolving Multilayer Networks}

\author{\IEEEauthorblockN{Alireza Hajibagheri}
\IEEEauthorblockA{University of Central Florida\\
alireza@eecs.ucf.edu}
\and
\IEEEauthorblockN{Gita Sukthankar}
\IEEEauthorblockA{University of Central Florida\\
gitars@eecs.ucf.edu}
\and
\IEEEauthorblockN{Kiran Lakkaraju}
\IEEEauthorblockA{Sandia National Labs\\
klakkar@sandia.gov}}

\maketitle

\begin{abstract}
\label{sec:abstract}
Networks extracted from social media platforms frequently include multiple types of links that dynamically change over time; these links can be used to represent dyadic interactions such as economic transactions, communications, and shared activities.  Organizing this data into a dynamic multiplex network, where each layer is composed of a single edge type linking the same underlying vertices, can reveal interesting cross-layer interaction patterns.  In coevolving networks, links in one layer result in an increased probability of other types of links forming between the same node pair.  Hence we believe that a holistic approach in which all the layers are simultaneously considered can outperform a factored approach in which link prediction is performed separately in each layer.  This paper introduces a comprehensive framework, \name/ (Multilayer Link Prediction), in which link existence likelihoods for the target layer are learned from the other network layers.  These likelihoods are used to reweight the output of a single layer link prediction method that uses rank aggregation to combine a set of topological metrics.  Our experiments show that our reweighting procedure outperforms other methods for fusing information across network layers.
\end{abstract}

\IEEEpeerreviewmaketitle

\section{Introduction}
\label{sec:intro}

As social media platforms offer customers more interaction options, such as \textit{friending},  \textit{following}, and \textit{recommending}, analyzing the rich tapestry of interdependent user interactions becomes increasingly complicated.  In this paper, we study two types of online societies: 1) players in a massively multiplayer online game (Travian)~\cite{hajibagheri2015conflict}  2) dialogs between Twitter users before, during, and after an exceptional event~\cite{omodei2015characterizing}.  Although standard social network analysis techniques~\cite{scott2012social} offer useful insights about these communities, there is relatively little theory from the social sciences on how to integrate information from multiple types of online interactions.

Rather than organizing this data into social networks separately chronicling the history of different forms of user interaction, dynamic multiplex networks~\cite{kivela2014} offer a richer formalism for modeling the social fabric of online societies.   A multiplex network is a multilayer network that shares the same set of vertices across all layers.  This network can be modeled as a graph $G=<V,E>$ where $V$ is the set of vertices and $E$ is the set of edges present in the graph. The dynamic graph $G=\{G_0,G_1,...,G_t\}$ represents the state of the network at different times.  The network is then defined as: $G_t = <V,E_t^1,...,E_t^M>$ with  $E_t^{\alpha} \subseteq V\times V$, $\forall \alpha \in \{1,...,M\}$, where each set $E_t^{\alpha}$ corresponds to the edge set of a distinct layer at time $t$.  Thus a dynamic multiplex network is well suited for representing diverse user activities over a period of time.

In this paper, we address the problem of predicting future user interactions from the history of past connections.  Assuming the data is represented as a graph, our goal is to predict the structure of graph $G_t$ using information from previous snapshots as well as other layers of the network.  Link prediction algorithms~\cite{hajibagheri2016link,liben2007link,menon2011link,scellato2011exploiting} have been implemented for many types of online social networks, including massively multiplayer online games and location-based social networks.
These systems offer great value to social networking services due to their practical applicability for friend recommendations and social network bootstrapping.   Although user profiles can be mined for additional data, topological approaches 1) perform well in many networks 2) preserve user privacy since they do not rely on actor information and 3) can be combined with node content approaches to enhance prediction performance.

Despite the fact that link prediction is a well studied problem, few link prediction techniques specifically address the problem of simultaneously predicting links across multiple networks~\cite{tang2012inferring,davis2013supervised,hristova2015multilayer}.  Basu et al.~\cite{basu2015} note that there are many real-world cases where interdependencies between processes cause the layers of a multiplex network to coevolve, resulting in a higher number of overlapping edges between the same node pair in different network layers.  In this paper, we explore the role of overlapping edges towards improving the performance of link prediction; our aim is to leverage the cross-layer link co-occurrence history to model coevolution in a multiplex network.

We propose a framework for multilayer link prediction (\name/) that integrates complementary information sources, including topological metrics, network dynamics, and overlapping edges.  \name/ uses a likelihood based method for learning cross-layer dependencies and a temporal decay function to model the network dynamics.  Rank aggregation is then employed to collect information from multiple topological metrics into one final scoring matrix. In the next section, we present related work on link prediction.  The proposed framework is described in Section~\ref{sec:framework}. Section~\ref{sec:experiments} presents a comparison of our method vs.\ two other approaches for fusing information across network layers.  We conclude in section~\ref{sec:conclusion} with a description of possible directions for future work.

\section{Related Work}
\label{sec:background}
A variety of computational approaches have been employed for predicting links in single layer networks, including supervised classifiers, statistical relational learning, matrix factorization, metric learning, and probabilistic graphical models (see surveys by \cite{lu2011link,al2011survey,zhang2014link} for a more comprehensive description).   Regardless of the computational framework, topological network measures are commonly used as features to describe node pairs and can be combined in a supervised or unsupervised fashion to do link prediction.~\cite{liben2007link}.  In this paper, we aggregate several of these metrics (listed in the next section),
but our framework can be easily generalized to include other types of features.

The primary focus of this paper is leveraging cross-layer information to improve link prediction in multiplex networks, although we also introduce our own single layer link prediction technique.  This process of using cross-layer information can be treated as a transfer learning problem where information is learned from a source network and applied to improve prediction performance the target network.  Tang et al.~\cite{tang2012inferring} introduced a transfer-based factor graph (TranFG) model which incorporates social theories into a semi supervised learning framework.  This model is then used to transfer supervised information from a source network to infer social ties in the target network.

Another strategy is to create more general versions of the topological measures that capture activity patterns in multiplex networks.  Davis et al.~\cite{davis2013supervised} introduced a probabilistically weighted extension of the Adamic/Adar measure for these networks. Weights are calculated by doing a triad census to estimate the probability of different link type combinations.  The extended Adamic/Adar metric is then used, along with other unsupervised link predictors, as input for a supervised classifier.  Similarly, Hristova et al.~\cite{hristova2015multilayer} extend the definition of network neighborhood by considering the union of neighbors across all layers.  These multilayer features are then combined in a supervised model to do link prediction.  One weakness with the above mentioned models is their inability to use temporal information accrued over many snapshots, rather than relying on a single previous snapshot.  In this paper, we evaluate two versions of our \name/ framework, a version that only uses topological metrics calculated from one time slice vs.\ multiple snapshots.

Conversely, there are a number of approaches that ignore cross-layer network dependencies, while using the history of changes between snapshots to predict future network dynamics.  We have experimented with two types of techniques: time series forecasting~\cite{hajibagheri2016link,soares2012time} and decay models~\cite{gao2011temporal}.  Soares and Prud\^{e}ncio~\cite{soares2012time} investigated the use of time series within both supervised and unsupervised link prediction frameworks. The core concept of their approach is that it is possible to predict the future values of topological metrics with time series; these values can either be used in an unsupervised fashion or combined in a supervised way with a classifier. In previous work, we introduced
a rate prediction model~\cite{hajibagheri2016link} that uses time series to predict the rate of link formation.  Our proposed framework, \name/, both models the rate of link formation in each layer and uses a decay model to account for changes in the topological metrics over time. In our results, we compare the improvements achieved by temporal vs. cross-layer modeling.

However, incorporating more features is not helpful, without an effective information fusion procedure.  Pujari et al.~\cite{pujari2012supervised} employed computational social choice algorithms for aggregating multiple topological features.  They evaluated the performance of two well-known rank aggregation methods, Borda and Kemeny, for single layer link prediction. In their method, weights are learned for each voter participating in the rank aggregation, where each topological metric is treated as a voter.   These weights are tuned to maximize the identification of positive examples or minimize negative examples.  
To extend their method to multiplex networks~\cite{pujari2015link}, the authors compute topological attributes for each network layer and combine them using 1) a simple aggregation of these scores across all layers or 2) an entropy-aggregation of values. These combinations are then used as a series of features in a decision tree model.  In this paper, we use rank aggregation to fuse our features and compare our procedure to their aggregation methods. Another example of a supervised framework that uses rank aggregation is  \textit{RankMerging}~\cite{tabourier2014rankmerging}. During a learning phase, weights are assigned to each unsupervised method using a training set of node pairs. The contribution of each ranking to the merged ranking is then computed using sliding indices.  At each step, the aim is to identify the ranking with the highest number of true predictions in the upcoming steps.  Rank aggregation methods can be highly effective, but the more complex social choice algorithms can suffer from high computational complexity, making them less effective for large datasets.  For this reason, we opted to use the Borda rank aggregation procedure in \name/.

\section{Node Similarity Metrics}
\label{sec:unsupervised}

This section provides a brief description of the topological and path-based metrics for encoding node similarity that are used within our \name/ framework to create ranked score lists for each node pair.  These techniques are often used in isolation as unsupervised methods for link prediction.  Note that $\Gamma(x)$ stands for the set of neighbors of vertex $x$ while $w(x,y)$ represents the weight assigned to the interaction between node $x$ and $y$.

\begin{itemize}
\item \textbf{Number of Common Neighbors (CN)}

The CN measure is defined as the number of nodes with direct relationships with both evaluated nodes $x$ and $y$~\cite{Newman01clusteringand}. For weighted networks, the CN measure is:

\begin{equation}\label{eq:CNW}
	CN(x,y)=\sum_{z\in|\Gamma(x)\cap\Gamma(y)|}w(x,z)+w(y,z)
\end{equation}

\item \textbf{Jaccard's Coefficient (JC)}

The JC measure assumes higher values for pairs of nodes who share a higher proportion of common neighbors relative to their total neighbors:

\begin{equation}\label{eq:JCW}
	JC(x,y) = \frac{\sum_{z \in \Gamma(x)\cap\Gamma(y)} w(x,z) + w(y,z)}{\sum_{a\in\Gamma(x)} w(x,a) + \sum_{b\in\Gamma(y)} w(y,b)}
\end{equation}

\item \textbf{Preferential Attachment (PA)}

The PA measure assumes that the probability that a new link originates from  node $x$ is proportional to its node degree. Consequently, nodes that already possess a high number of relationships tend to create more links ~\cite{barabasi2009scale}: 

\begin{equation}\label{eq:PAW}
	PA(x,y)=\sum_{z_{1}\in\Gamma(x)} w(x,z_1) \times \sum_{z_{2}\in\Gamma(y)} w(y,z_2)
\end{equation}
\newpage
\item \textbf{Adamic-Adar Coefficient (AA)}

This metric~\cite{adamic2003friends} is closely related to Jaccard's coefficient in that it assigns a greater importance to common neighbors who have fewer neighbors. Hence, it measures the exclusivity of the relationship between a common neighbor and the evaluated pair of nodes:

\begin{equation}\label{eq:AAW}
	AA(x,y)=\sum_{z\in\Gamma(x)\cap\Gamma(y)}\frac{w(x,z) + w(y,z)}{log(1 + \sum_{c\in\Gamma(z)} w(z,c))}
\end{equation}

\item \textbf{Resource Allocation (RA)}
	
RA was first proposed in~\cite{zhou2009predicting} and is based on physical processes of resource allocation:

\begin{equation}\label{eq:RAW}
	RA(x,y)=\sum_{z\in\Gamma(x)\cap\Gamma(y)}\frac{w(x,z) + w(y,z)}{\sum_{c\in\Gamma(z)} w(z,c)}
\end{equation}

\item \textbf{Page Rank (PR)} 
	
The PageRank algorithm~\cite{brin2012reprint} measures the significance of a node based on the significance of its neighbors.  We use the weighted PageRank algorithm proposed in\cite{ding2011applying}:

\begin{equation}\label{eq:PRW}
	PR_{w}(x)= \alpha \sum_{k\in\Gamma(x)} \frac{PR_{w}(x)}{L(k)} + (1-\alpha) \frac{w(x)}{\sum_{y=1}^{N} w(y)}
\end{equation}

where $L(x)$ is the sum of outgoing link weights from node $x$, and $\sum_{y=1}^{N} w(y)$ is the total weight across the whole network.

\item \textbf{Inverse Path Distance (IPD)}

The Path Distance measure for unweighted networks simply counts the number of nodes along the shortest path between $x$ and $y$ in the graph. Note that $PD(x,y) = 1$ if two nodes $x$ and $y$ share at least one common neighbor. In this article, the Inverse Path Distance is used to measure the proximity between two nodes, where:

\begin{equation}\label{eq:IPDW}
IPD(x,y) = \frac{1}{PD(x,y)}
\end{equation}

IPD is based on the intuition that nearby nodes are likely to be connected. In a weighted network, IPD is defined by the inverse of the shortest weighted distance between two nodes. 

\item \textbf{Product of Clustering Coefficient (PCF)} 

The clustering coefficient of a vertex $v$ is defined as:

\begin{equation}
	PCF(v) = \frac{3 \times \mbox{\# of triangles adjacent to v}}{\mbox{\# of possible triples adjacent to v}}
\end{equation}

To compute a score for link prediction between the vertex $x$ and $y$, one can multiply the clustering coefficient score of $x$ and $y$.
\end{itemize}
Section~\ref{sec:experiments} compares \name/ vs.\ unsupervised versions of these approaches.

\section{Proposed Method}
\label{sec:framework}

\name/ is a hybrid architecture that utilizes multiple components to address different aspects of the link prediction task.  We seek to extract information from all layers of the network for the purpose of link prediction within a specific layer known as the target layer. To do so, we create a weighted version of the original target layer where interactions and connections that exist in other layers receive higher weights. After reweighting the layer, we employ the collection of node similarity metrics described in the previous section on the weighted network.  To express the temporal dynamics of the network, we use a decay model on the time series of similarity metrics to predict future values.  Finally, the Borda rank aggregation method is employed to combine the ranked lists of node pairs into a single list that predicts links for the next snapshot of the target network layer.  Each component of the model is explained in more detail in the following sections.

\subsection{Multilayer Likelihood Assignment and Edge Weighting}
This component leverages information about cross-layer link co-occurrences.  During the coevolution process, links may be engendered due to activity in other network layers.  Some layers may evolve largely independently of the rest of the network, whereas links in other layers may be highly predictive of links in the target layer.
In our proposed method, a weight is assigned to each layer based on its influence on the target layer. Weights are calculated using a likelihood function:
\begin{equation}
	w_i = Likelihood(\text{Link in }L^{Target} | \text{Link in }L^i)
\end{equation}
\noindent
where $L^i$ and $w_i$ represent the $i$th layer and the weight calculated for it respectively. $L^{Target}$ indicates the target layer for which we want to predict future links. The \textit{Likelihood} function computes the similarity between the target layer and the $i$th layer; to do this, we use the current ratio of overlapping edges.  Next, we calculate weights for every node pair by checking the link correspondence between two layers using the likelihood of a link being present in the target layer given the existence of the link in the other layer at any other previous snapshot. This orders other layers in terms of their relative importance for a specific target layer. The process assigns higher weights to node pairs which occur in more than one layer (multiplex edges).  The rate of link formation is incorporated into the model as the first term of the edge weight.
Algorithm~\ref{alg:weighting} shows the process of assigning likelihoods to layers and reweighting the adjacency matrix.

\begin{algorithm}
    \caption{Likelihood Assignment and Edge Weighting}
    \label{alg:weighting}
    \begin{algorithmic}[1] 
\STATE Input: Edge sets $(E^{1},..., E^{M})$ for $M$ layers where $E^{\alpha}$ is the edge set of target layer   
\STATE Output: $E_{w}^{\alpha}$ weighted adjacency matrix for layer $\alpha$ (target layer)
\\ //Calculate weights for the layers
\FOR{$i \in \{1,2,...,M\}-\{\alpha\}$}
\STATE $w_i = Likelihood(\text{Link in }L^{\alpha} | \text{Link in }L^i)$
\ENDFOR
\\ //Weighting target layer
\FOR{edge $e \in E^{\alpha}$}
\STATE $w_e = rate + \sum_{i=1 \& i \neq \alpha}^{M} w_i \times linkExist(e)$
\ENDFOR
  \end{algorithmic}
\end{algorithm}

The term \textit{rate} is defined as the average value of the source node's out-degree over previous timesteps. Function \textit{linkExist} is used to obtain information about a link's existence in other layers during previous snapshots. It checks each layer for the presence of an edge and returns 1 if an edge is present in that layer.

\subsection{Temporal Link Structure}
Given the network history for $T$ time periods, we need to capture the temporal dependencies of the coevolution process.  To do so, our framework uses a weighted exponentially decaying model~\cite{acar2009link}. Let $\{Sim_t(i, j),t = t_0+1, ..., t_0+T\}$ be a time series of similarity score matrices generated by a node similarity metric on a sliding window of $T$ successive temporal slices. An aggregated weighted similarity matrix is constructed as follows:

\begin{equation}
Sim_{(t_0+1)\sim(t_0+T)}(i,j) = \sum_{t=t_0+1}^{t_0+T} \theta^{t_0+T-t} Sim_t(i,j)
\end{equation}
\noindent where the parameter $\theta \in [0, 1]$ is the smoothing weight for previous time periods. Different values of $\theta$ modify the importance assigned to the most or least recent snapshots before current time $t+1$. This procedure generates a composite temporal score matrix for every node similarity metric.  $Sim_{(t_0+1)\sim(t_0+T)}$ (shortened to $Sim$) is used by the algorithm as a summary of network activity, encapsulating the temporal evolution of the similarity matrix.

\subsection{Rank Aggregation}

Before describing the final step of our approach, let us briefly discuss existing methods for \textit{ranked list aggregation/rank aggregation}. List merging or list aggregation refers to the process of combining a number of lists with the same or different numbers of elements in order to get one final list including all the elements. In rank aggregation, the order or rank of elements in input lists is also taken into consideration. The input lists can be categorized as \textit{full, partial, or disjoint lists}. Full lists contain exactly the same elements but with a different ordering, partial lists may have some of the elements in common but not all, and disjoint lists have completely different elements. In this case, we are only dealing with full lists since each similarity metric produces a complete list for the same set of pairs, differing only in ordering. 

Several rank aggregation methods are described in~\cite{sculley2007rank}, including Borda's, Markov chain, and median rank methods. Borda's method is a \textit{rank-then-combine} method originally proposed to obtain a consensus from a voting system. Since it is based on the absolute positioning of the rank elements and not their relative rankings, it can be considered a truly positional method. For every element in the lists, a Borda score is calculated and elements are ranked according to this score in the aggregated list. For a set of complete ranked lists $L = [L_1, L_2, L_3, ....,L_k]$, the Borda score for an element $i$ and a list $L_k$ is given by:
\begin{equation}
B_{L_k}(i) = \{count(j) | L_k(j) < L_k(i) \& j\in L_k\}
\end{equation}
\noindent
The total Borda score for an element is given as:
\begin{equation}
B(i) = \sum_{t=1}^{k} B_{L_t(i)}
\end{equation}
\begin{figure*}
  \centering 
\subfloat[]{\includegraphics[width=0.2\textwidth]{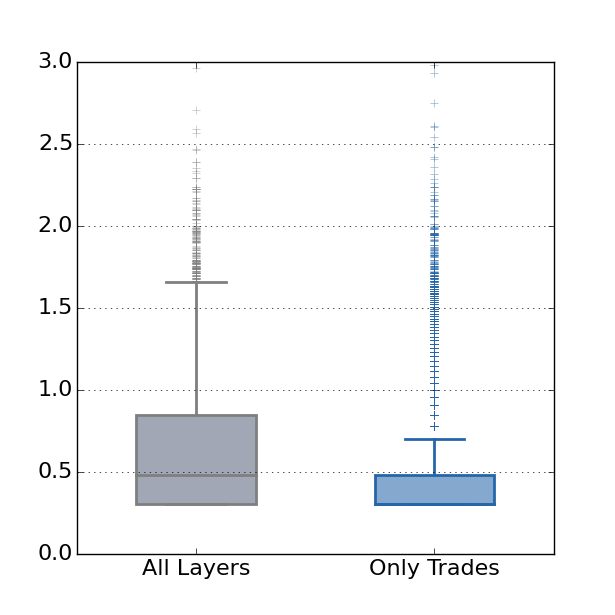}\label{trades}}
\subfloat[]{\includegraphics[width=0.2\textwidth]{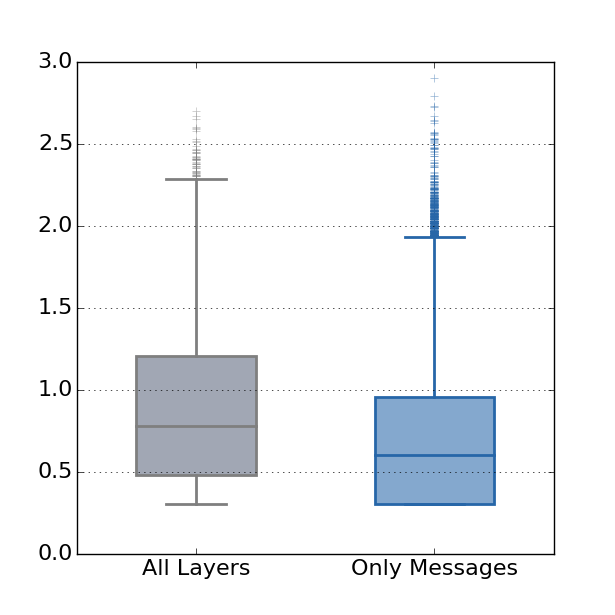}\label{messages}}
\subfloat[]{\includegraphics[width=0.2\textwidth]{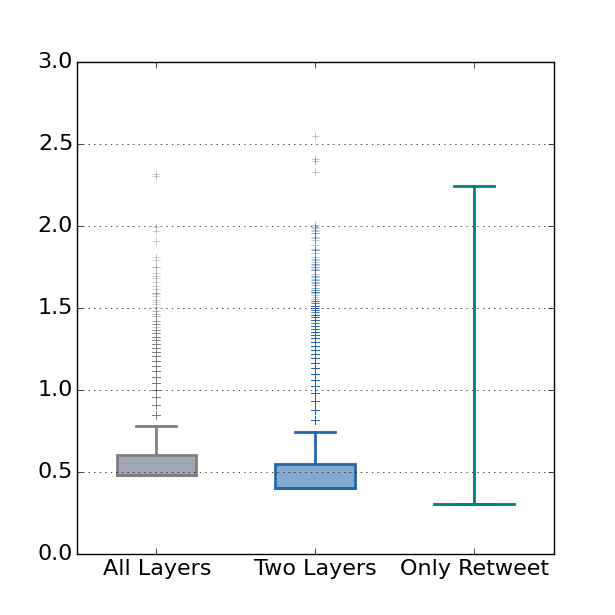}\label{retweets}}
\subfloat[]{\includegraphics[width=0.2\textwidth]{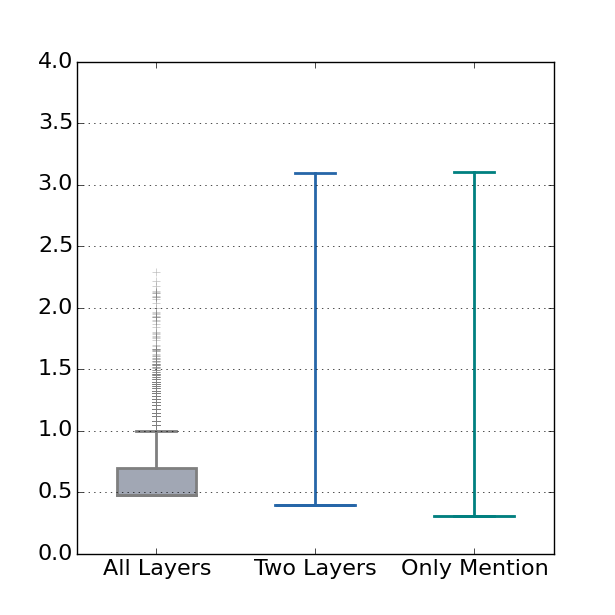}\label{mentions}}
\subfloat[]{\includegraphics[width=0.2\textwidth]{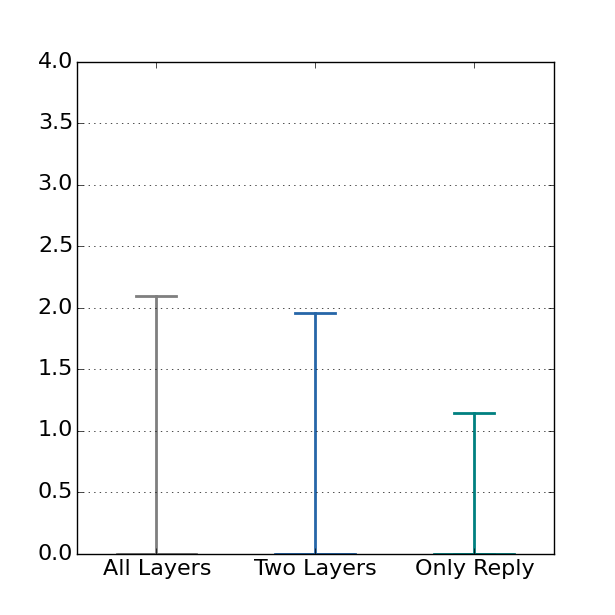}\label{replys}}
    \caption{Log scale box-whisker plots for user interactions in different layers of the network: (a) Travian (Trades) (b) Travian (Messages) (c) Cannes2013 (Retweets) (d) Cannes2013 (Mentions) (e) Cannes2013 (Replies)}
    \label{fig:interactions}
\end{figure*}
\begin{figure}
  \centering 
\includegraphics[width=0.45\textwidth]{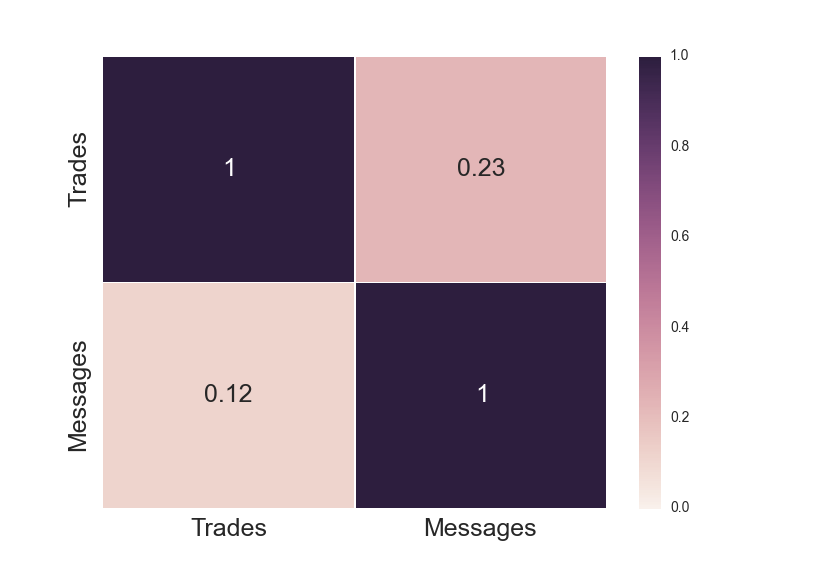}\label{travian}
\includegraphics[width=0.45\textwidth]{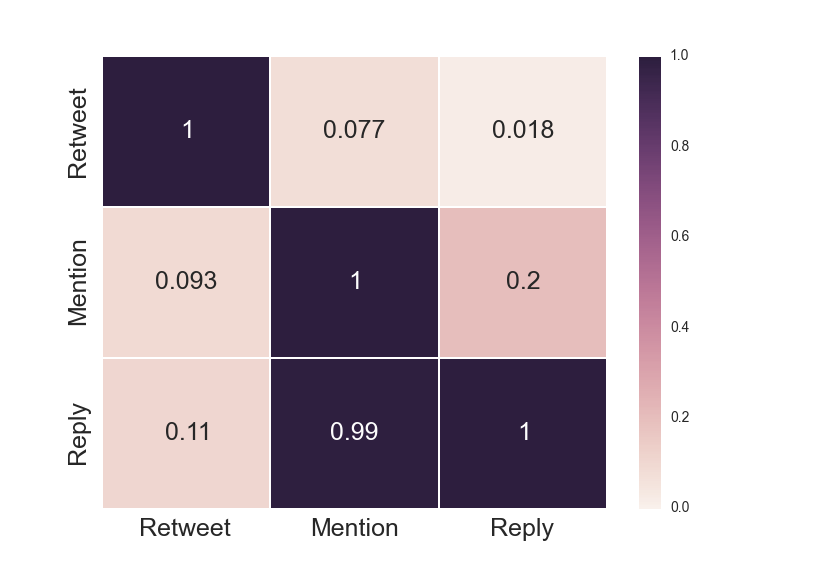}\label{cannes}
    \caption{Heatmap representing the edge overlap between pairs of layers for datasets (a) Travian (b) Cannes2013}
    \label{fig:overlap_figures}
\end{figure}
\noindent
Borda's method is computationally cheap, which is a highly desirable property for link prediction in large networks.  

Algorithm~\ref{alg:framework} shows our proposed framework which incorporates edge weighting, the temporal decay model, and rank aggregation to produce an accurate prediction of future links in a dynamic multilayer network. The Borda function produces the final output of the \name/ framework. Results of the proposed algorithm are compared with other state-of-the-art techniques in the next section.







\begin{algorithm}
    \caption{Multilayer Link Prediction Framework (\name/)}
    \label{alg:framework}
    \begin{algorithmic}[1] 
\STATE Input: Weighted edge sets of the target layer for $T$ previous snapshots 
\STATE Output: Temporal aggregated score matrix $S$ for the target layer
\FOR{each node similarity metric $u$}
\FOR{$t \in \{1,...,T\}$}
\STATE Calculate score matrix $Sim_{t0+t}^u$
\ENDFOR
\STATE Calculate temporal similarity matrix $Sim^u$
\ENDFOR
\STATE Final score matrix $S = Borda(Sim^1,...,Sim^u)$
  \end{algorithmic}
\end{algorithm}

\section{Experimental Study}
\label{sec:experiments}

This paper evaluates the \name/ framework on networks extracted from two real-world datasets, Travian and Cannes2013.  To investigate the impact of each component of our proposed method, not only do we compare our results with two other approaches for fusing cross-layer information, but we also analyze the performance of ablated versions of our method.  The complete method, MLP (Hybrid), is compared with MLP (Decay Model + Rank Aggregation) and MLP (Weighted + Rank Aggregation).  All of the algorithms were implemented in Python and executed on a machine with the Intel(R) Core i7 CPU and 24GB of RAM for the purpose of fair comparison.  Our implementation uses Apache Spark to speed the link prediction process.

\subsection{Datasets}
We use two real-world dynamic multilayer networks to demonstrate the performance of our proposed algorithm. These networks are considerably disparate in structure and were selected from different domains (a massively multiplayer online game (MMOG) and an event-based Twitter dataset).  Table~\ref{tab:network_stats} provides the network statistics for each of the datasets:
\begin{itemize}

\item \textbf{Travian MMOG}~\cite{hajibagheri2015conflict}  Travian is a browser-based, real-time strategy game in which the players compete to create the first civilization capable of constructing a Wonder of the World. The experiments in this paper were conducted on a 30 day period in the middle of the Travian game cycle.  In Travian, players can execute different game actions including: sending messages, trading resources, joining alliances, and attacking enemy villages. In this research, we focus on networks created from trades and messages. 
	
\item \textbf{Twitter Interactions}~\cite{omodei2015characterizing} This dataset consists of Twitter activity before, during, and after an ``exceptional'' event as characterized by the volume of communications.  Unlike most Twitter datasets which are built from follower-followee relationships, links in this multilayer network correspond to retweeting, mentioning, and replying to other users. The Cannes2013 dataset was created from tweets about the Cannes film festival that occurred between May 6,2013 to June 3, 2013. Each day is treated as a separate network snapshot.
\end{itemize}

\begin{table}
\caption{Dataset Summary: Number of edges, nodes, and snapshots for each network layer}
\label{tab:network_stats}
\begin{center}
\begin{tabular}{llclccc}
\textbf{Dataset} & & 
\textbf{Travian} & & 
\textbf{Cannes2013} & &
 \\
\midrule
\textbf{No. of Nodes} & & 2,809 & &438,537& \\
\textbf{No. of Snapshots} & & 30 & & 29 &
\\
\midrule
\textbf{Layers/No. of Edges} 
& Trades & 87,418 & Retweet & 496,982 & & \\
& Messages & 44,956 & Mention & 411,338 & & \\
&  &  & Reply & 83,534 & & \\

\bottomrule
\end{tabular}
\end{center}
\end{table}

\subsection{Evaluation Metrics}
For the evaluation, we measure receiver operating characteristic (ROC)  curves for the different approaches. The ROC curve is a plot of the \textit{true positive rate (tpr)} against the \textit{false positive rate (fpr)}. These curves show achievable true positive rates (TP) with respect to all false positive rates (FP) by varying the decision threshold on probability estimations or scores. For all of our experiments, we report area under the ROC curve (AUROC), the scalar measure of the performance over all thresholds. Since link prediction is highly imbalanced, straightforward accuracy measures are well known to be misleading; for example, in a sparse network, the trivial classifier that labels all samples as missing links can have a 99.99\% accuracy. 



\subsection{Analysis of Cross-layer Interaction}
Figure~\ref{fig:interactions} shows log scale box-whisker plots that depict the frequency of interactions between users who are connected across multiple layers.  We compare the frequency of interactions in cases where the node pair is connected on all layers vs.\ the frequency of being connected in a single layer (Travian) or less than all layers (for Cannes which has three layers).  As expected, in cases where users are connected on all layers, the number of interactions (trades, messages, retweets, mentions and replies) is higher. The heatmap of the number of overlapping edges between different network layers (Figure~\ref{fig:overlap_figures}) suggests that a noticeable  number of edges are shared between all layers. This clearly indicates the potential value of cross-layer information for the link prediction task on these datasets.  Our proposed likelihood weighting method effectively captures the information revealed by our analysis.

\begin{table*}[t]
\caption{AUROC performances for a target layer averaged over all snapshots with a sliding time window of $T=3$ for Travian layers and $T=5$ for Cannes2013 layers used in the decay model.  Variants of our proposed framework are shown at the top of the table, followed by standard unsupervised methods.  The algorithms shown in the bottom half of the table are techniques for multiplex networks proposed by other research groups.  The best performer is marked in bold.}
	
\label{tab:auroc}
\begin{center}
  \bgroup
  \def\arraystretch{1.5}
  \begin{tabular*}{1\textwidth}{p{4.5cm}C{2.4cm}cccc}
  	\hline
	Algorithms / Networks & Travian (Trade) & Travian (Message) & Cannes2013 (Retweet) & Cannes2013 (Mention) & Cannes2013 (Reply)\\ \hline
\textbf{\name/ (Hybrid)} &\textbf{0.8209$\pm$0.0016}&\textbf{0.8036$\pm$0.0023}&\textbf{0.8116$\pm$0.0025}&\textbf{0.8345$\pm$0.0035}&\textbf{0.8393$\pm$0.0025} \\
\textbf{\name/ (Likelihood + Rank Aggregation)} &0.8024$\pm$0.0013&0.7906$\pm$0.0021&0.8088$\pm$0.0027&0.8141$\pm$0.0036&0.8160$\pm$0.0029\\
\textbf{\name/ (Decay Model + Rank Aggregation)} &0.7226$\pm$0.0029&0.7310$\pm$0.0021&0.7267$\pm$0.0025&0.7283$\pm$0.0028&0.7331$\pm$0.0018\\
\textbf{Likelihood} &0.7702$\pm$0.0332&0.7606$\pm$0.0410&0.7524$\pm$0.0218&0.7806$\pm$0.0517&0.7573$\pm$0.0423\\
\textbf{Rank Aggregation} &0.6941$\pm$0.0010&0.7120$\pm$0.0014&0.7001$\pm$0.0020&0.7062$\pm$0.0023&0.7004$\pm$0.0027\\
    \hline
\textbf{Common Neighbors} &0.6565$\pm$0.0024&0.6671$\pm$0.0021&0.6999$\pm$0.0019&0.7050$\pm$0.0031&0.6999$\pm$0.0012\\
\textbf{Jaccard Coefficient} &0.6287$\pm$0.0023&0.6803$\pm$0.0031&0.5944$\pm$0.0016&0.7331$\pm$0.0016&0.7107$\pm$0.0029\\
\textbf{Preferential Attachment} &0.7094$\pm$0.0019&0.6374$\pm$0.0013&0.5846$\pm$0.0019&0.6122$\pm$0.0020&0.5869$\pm$0.0026\\
\textbf{Adamic/Adar} &0.6354$\pm$0.0034&0.7002$\pm$0.0028&0.6999$\pm$0.0026&0.6419$\pm$0.0022& 0.5160$\pm$0.0028\\
\textbf{Resource Allocation} &0.6254$\pm$0.0052&0.6902$\pm$0.0028&0.5973$\pm$0.0021&0.6222$\pm$0.0019&0.6717$\pm$0.0036\\
\textbf{Page Rank} &0.5954$\pm$0.0016&0.6871$\pm$0.0021&0.6604$\pm$0.0024&0.6301$\pm$0.0032&0.6131$\pm$0.0017\\
\textbf{Inverse Path Distance} &0.5723$\pm$0.0037&0.6506$\pm$0.0032&0.6312$\pm$0.0030&0.6410$\pm$0.0020&0.5614$\pm$0.0036\\
\textbf{Clustering Coefficient} &0.5804$\pm$0.0023&0.6332$\pm$0.0025&0.5702$\pm$0.0201&0.6213$\pm$0.0105&0.5220$\pm$0.0041\\ 
\hline
\textbf{Average Aggregation} 
&0.7446$\pm$0.0300&0.7521$\pm$0.0201&0.7405$\pm$0.0026&0.7366$\pm$0.0107&0.7611$\pm$0.0035\\
\textbf{Entropy Aggregation} 
&0.7310$\pm$0.0038&0.7630$\pm$0.0201&0.7515$\pm$0.0030&0.7584$\pm$0.0311&0.7441$\pm$0.0022\\
\textbf{Multilayer Common Neighbors} &0.7293$\pm$0.0040&0.6429$\pm$0.0130&0.6723$\pm$0.0028&0.7163$\pm$0.0027&0.7328$\pm$0.0019\\
\textbf{Multilayer Jaccard Coefficient} 
&0.6663$\pm$0.0312&0.6193$\pm$0.0124&0.5803$\pm$0.0031&0.7360$\pm$0.0022&0.7219$\pm$0.0016\\
\textbf{Multilayer Preferential Attachment} 
&0.7226$\pm$0.0098&0.6459$\pm$0.0127&0.5801$\pm$0.0034&0.6403$\pm$0.0031&0.6212$\pm$0.0028\\
\textbf{Multilayer Adamic/Adar} 
&0.6712$\pm$0.0104&0.6899$\pm$0.0310&0.6709$\pm$0.0036&0.6687$\pm$0.0026& 0.5523$\pm$0.0033\\
     
    \hline
  \end{tabular*}
  \egroup
 	\end{center}
\end{table*}

\subsection{Performance of Multilayer Link Prediction} 

For our experiments, we adopted a moving-window approach to evaluate the performance of our temporal multilayer link prediction algorithm. Given a specified window size $T$, for each time period $t (t > T)$, graphs of $T$ previous periods $(G_{t−T},...,G_{t−1})$ (where each graph consists of $M$ layers) are used to predict links that occur at the target layer $\alpha$ in the current period $(G_t^{\alpha})$. To assess our proposed framework and study the impact of its components, we compare against the following baselines:

\begin{itemize}
\item \textbf{\name/ (Hybrid)}:  incorporates all elements discussed in the framework section. It utilizes the likelihood assignment and edge weighting procedure to extract cross-layer information.  Node similarity scores are modified using the temporal decay model and combined with Borda rank aggregation. 
\item \textbf{\name/ (Likelihood + Rank Aggregation)}: This method only uses the aggregated scores calculated from the graphs weighted with cross-layer information. It does not consider the temporal aspects of network coevolution.
\item \textbf{\name/ (Decay Model + Rank Aggregation)}: This method does not use the cross-layer weighting scheme and relies on temporal information alone to predict future links. The final aggregated score matrix is calculated based on forecast values at time $t$ for each node similarity metric using the decay model.
\item \textbf{Likelihood}: Weights generated by the cross-layer likelihood assignment procedure are treated as scores for every node pair. We then sort the pairs based on their score and calculate the AUROC.
\item \textbf{Rank Aggregation}: This method is a simple aggregated version of all unsupervised scoring methods using the Borda's rank aggregation method applied to node similarity metrics from the target layer.
\item \textbf{Unsupervised Methods}: The performance of our proposed framework is compared with eight well-known unsupervised link prediction methods described in Section~\ref{sec:unsupervised}. All unsupervised methods are applied to the binary static graph from time $0$ to $t-1$ in order to predict links at time $t$. Only the structure of the target layer is used.
\item \textbf{Average Aggregation}: In order to extend the rank aggregation model to include information from other layers of the network, we use the idea proposed in~\cite{pujari2015link}.  Node similarity metrics are aggregated across all layers. So for attribute $X$ (Common Neighbors, Adamic/Adar, etc.) over $M$ layers the following is defined:
\begin{equation}
X(u,v) = \frac{\sum_{\alpha=1}^M X(u,v)^{\alpha}}{M}
\end{equation}
where $X(u,v)$ is the average score for nodes $u$ and $v$ across all layers and $X(u,v)^{\alpha}$ is the score at layer $\alpha$. Borda's rank aggregation is then applied to the extended attributes to calculate the final scoring matrix.
\item \textbf{Entropy Aggregation}: Entropy aggregation is another extended rank aggregation model proposed in~\cite{pujari2015link} where $X(u,v)$ is defined as follows:
\begin{equation}
X(u,v) = -\sum_{\alpha=1}^M \frac{X(u,v)^{\alpha}}{X_{total}} \log(\frac{X(u,v)^{\alpha}}{X_{total}})
\end{equation}
where $X_{total} = \sum_{\alpha=1}^M X(u,v)^{\alpha}$. The entropy based attributes are more suitable for capturing the distribution of the attribute value over all dimensions. A higher value indicates a uniform distribution of attribute values across the multiplex layers.
\item \textbf{Multilayer Unsupervised Methods}: Finally, using the definition of core neighborhood proposed in~\cite{hristova2015multilayer}, we extend four unsupervised methods (Common Neighbors, Preferential Attachment, Jaccard Coefficient and Adamic/Adar) to their multilayer versions.
\end{itemize}

Table~\ref{tab:auroc} shows the results of different algorithms on the Travian and Cannes2013 datasets. With 30 days of data from Travian and 27 days for Cannes2013, we were able to extensively compare the performance of the proposed methods and the impact of using different elements. Bold numbers indicate the best results on each target layer considered; \name/ (Hybrid) is the best performing algorithm in all cases.


\section{Discussion}
In this section, we discuss the most interesting findings:

\textbf{Does rank aggregation improve the performance of the unsupervised metrics?} As shown in Table~\ref{tab:auroc}, although the aggregated scores matrix produced by Borda's method achieves better results than unsupervised methods in some cases (Travian message, Cannes2013 retweet and mention networks) and comparable results on others (Travian trade and Cannes2013 reply networks), it is not able to significantly outperform all unsupervised methods in any of the networks. As discussed before, we are using the simple Borda method for the rank aggregation which does not consider the effect of each ranker on the final performance. While adding weights to the rankers or using more complex rank aggregation models such as Kemeny might achieve better results, it has been shown that those approaches have high computational complexity which makes them less suitable for large real-world networks~\cite{pujari2012supervised,tabourier2014rankmerging}. Despite the fact that the rank aggregation alone
does not significantly improve the overall performance of the link prediction task, it enables us to effectively fuse different kinds of information (edge and node features, nodes similarity, etc.).  

On the other hand, the Average and Entropy Aggregation methods which are designed to consider attribute values from other layers are able to outperform regular Rank Aggregation and \name/ (Decay Model + Rank Aggregation). However, both methods use the static structure of all snapshots from time $0$ to $t-1$, while \name/ (Decay Model + Rank Aggregation) only incorporates the past $T$ snapshots which makes it more suitable for large networks.

\textbf{Does the likelihood assignment procedure outperform the unsupervised scores?} To study the ability of our likelihood weighting method to model the link formation process, we generate results for two methods: using likelihood explicitly as a scoring method as well as using the values to generate a weighted version of the networks. First, the \textit{Likelihood} method is used in isolation to demonstrate the prediction power of its weights as a new scoring approach. Table~\ref{tab:auroc} shows significant improvements on unsupervised scores as well as the aggregated version of them. As expected, the more overlap between the target layer and predictor layers, the more performance improvement \textit{Likelihood} achieves. As an example, Likelihood achieves $\sim7\%$ of improvement on Travian (Trade) compared with $\sim 5\%$ of improvement on Travian (Message). Not only is there a lower rate of overlapping edges between those layers, but also the number of interactions is higher than the two other layers. The same holds true for Cannes2013 (Retweet) compared with the mention and reply layers.

On the other hand, the method introduced in Algorithm~\ref{alg:weighting} generates a weighted version of input graphs which is used to generate a weighted version of unsupervised methods to produce the final scoring matrix. This paired with the rank aggregation method generates significantly better average AUROC performance compared with other proposed methods. Also, when temporal information from previous snapshots of the network is included, \name/ (Hybrid) outperforms other variants of \name/ as well as well-known unsupervised methods. This indicates the power of overlapping links in improving the performance of link prediction in coevolving multilayer networks.

\textbf{Does including temporal information improve AUROC performance?} The importance of incorporating temporal information into link prediction has been thoroughly discussed in our previous work~\cite{hajibagheri2016link}. However, here we are interested in analyzing the impact of this information on improving the performance of MLP.  For that purpose, first, the decay model is employed in \name/ (Decay Model + Rank Aggregation) to determine whether it improves the results generated by the aggregated score matrix. The final aggregated score matrix is calculated based on forecast values at time $t$ for each unsupervised method using the decay model. As expected, this version of \name/ is able to achieve up to $\sim3\%$ of AUROC improvement using only information from the last three and five snapshots of the Travian and Cannes2013 networks respectively. On the other hand, we observed the same pattern when the decay model was added to \name/ (Hybrid) along with likelihood and rank aggregation. Link prediction using the scores generated by our hybrid method outperformed all other proposed and existing methods. The results presented here have been obtained using $T=3$ for the Travian dataset and $T=5$ for Cannes2013. These values are based on experiments performed on both datasets. While for Travian layers, increasing the value of $T$ tends to improve the prediction performance slightly until $T=3$; higher values of $T$ may decrease the performance. The same pattern occurs for Cannes2013 layers when $T=5$. Similarly, the value of $\theta$ is set to $0.4$ for both datasets.

In summary, \name/ (Decay Model + Rank Aggregation) is able to achieve results comparable to other baseline methods except Average and Entropy Aggregation since they benefit from the entire graph structure. Although rank aggregation by itself is not able to significantly improve the performance of unsupervised methods, paired with decay models and taking temporal aspects of the network, it can achieve better performance. On the other hand, the multilayer versions of the neighborhood based unsupervised methods are able to improve average AUROC performance, however the results are inconsistent and they achieve lower performance in many cases. Finally, both \name/ (Hybrid) and \name/ (Likelihood + Rank Aggregation) achieve higher performance compared with all other methods illustrating the importance of the cross-layer information created by the network coevolution process.
A paired two-sample $t$-test is used to indicate the significance of the results produced by each method where the $p$-value is smaller than $0.0001$. It is worth mentioning that, even though \name/ (Hybrid) is able to outperform all other methods, its performance is not significantly better than \name/ (Likelihood + Rank Aggregation) in the case of Travian (Message) and Cannes (Retweet). 



\section{Conclusion and Future Work}
\label{sec:conclusion}
In this paper, we introduce a new link prediction framework, \name/ (Multilayer Link Prediction), that employs a holistic approach to accurately predict links in dynamic multiplex networks using a collection of topological metrics, the temporal patterns of link formation, and overlapping edges created by network coevolution.
Our analysis on real-world networks created by a variety of social processes suggests that \name/ effectively models multiplex network coevolution in many domains.

The version of Borda's method used in this research assigns the same weight to all rankers. However, for different networks, each scoring method might add differing value to the final scoring matrix. In future work, it would be interesting to use weighted Borda to calculate final scores. Also, while using more network features often increases the performance of a link prediction algorithm, this might not be true for all networks. Thus it may be useful to employ a feature selection algorithm to identify the best subset of unsupervised methods to be used in \name/, based on performance improvements in early snapshots.

\section*{Acknowledgments}
Research at University of Central Florida was supported by NSF award IIS-0845159. Sandia National Laboratories is a multi-program laboratory managed and operated by Sandia Corporation, a wholly owned subsidiary of Lockheed Martin Corporation, for the U.S. Department of Energy's National Nuclear Security Administration under contract DE-AC04-94AL85000.

\end{document}